\documentclass[9pt,twoside]{extarticle}

\usepackage[margin=0.75in, top=0.75in, bottom=0.75in,
            columnsep=0.25in]{geometry}
\usepackage{multicol}

\usepackage[T1]{fontenc}
\usepackage{mathptmx}
\usepackage[scaled=0.86]{helvet}
\usepackage{microtype}

\usepackage{amsmath,amssymb,bm}
\usepackage{mathtools}

\usepackage{float}

\usepackage{graphicx}
\usepackage{booktabs}
\usepackage{dcolumn}
\usepackage[rightcaption]{sidecap}
\usepackage[labelfont=bf, 
            font=small,
            labelsep=period]{caption}
\usepackage{enumitem}

\usepackage{xcolor}
\usepackage{url}
\newcommand{\doi}[1]{\href{https://doi.org/#1}{\detokenize{#1}}}
\usepackage{hyperref}
\usepackage{hycolor}
\hypersetup{colorlinks=true,citecolor=[rgb]{0,0.18,0.68}, 
urlcolor=[rgb]{0,0.18,0.68}, linkcolor=[rgb]{0,0.18,0.68}, final = true}
\renewcommand{\url}[1]{\href{#1}{\detokenize{#1}}}
\usepackage{relsize}

\usepackage[super,compress,sort]{natbib}
\setcitestyle{numbers,super,open={},close={}}

\usepackage{titlesec}
\titleformat{\section}{\small\bfseries\sffamily\uppercase}{\thesection.}{0.5em}{}
\titleformat{\subsection}{\small\bfseries\sffamily}{\thesubsection.}{0.5em}{}
\titleformat{\subsubsection}{\small\itshape}{\thesubsubsection.}{0.5em}{}

\usepackage{pdfpages}

\usepackage{fancyhdr}
\usepackage{abstract}

\usepackage{siunitx}
\titleformat{\section}
    {\fontsize{12}{10}\bfseries\sffamily\uppercase}
    {\thesection.}{0.5em}{}

  \def\be{\begin{equation*}}
  \def\ee{\end{equation*}}
  \def\ba{\begin{eqnarray}}
  \def\ea{\end{eqnarray}}
  
  \def\fref#1{Fig.~\ref{#1}}

  \def\bt{\textrm} 
  
  \def\nsb#1{\noindent\textbf{\bt{#1~}}}
  \def\nsi#1{\noindent\textit{\bt{#1~--}}}
  \definecolor{or}{RGB}{234,142,53}
  \definecolor{gr}{RGB}{150,150,150}
  \definecolor{bl}{RGB}{54,152,187}

  \newcommand{\ie}{\textit{i.e.}}
  \newcommand{\eg}{\textit{e.g.}}

  \definecolor{YKB}{rgb}{0.00,0.18,0.65}

\begin{document}

\title{\LARGE\bfseries\sffamily What is a Musical Scale? Regularity
       and Convention in the Organization of Pitch.}

\author{
    John M. McBride\textsuperscript{1,2,*} 
}

\date{
    \small
    \textsuperscript{1}Department of Behavioral and Cognitive Biology, 
    University of Vienna, Vienna, Austria\\[0.5em]
    \textsuperscript{2}Acoustics Research Institute,
    Austrian Academy of Science, Vienna, Austria\\[0.5em]
    \textsuperscript{*}Correspondence: 
    \href{mailto:jmmcbride@protonmail.com}{jmmcbride@protonmail.com}\\
}

\twocolumn[
  \begin{@twocolumnfalse}
    \maketitle
  \end{@twocolumnfalse}
]

\begin{abstract}
  Musical scales are near-universal in human music, and most
  readers will feel they already know what a scale is. On closer
  inspection, however, the literature lacks a consensus definition: which
  conditions are necessary and sufficient shifts across
  disciplines and traditions, and the term turns out to cover
  several distinct objects. I argue this is less a failure of
  rigour than a sign that ``scale'' names several related objects:
  prescriptive abstractions, instrument tunings, statistical
  regularities in performed pitch, perceptual categories, social conventions.
  I adopt an empirical definition -- a scale as a statistical
  regularity in pitch organisation relative to a tonic -- that is
  portable across traditions and computable from recordings, and
  situate it alongside the other senses of the term. Even this
  empirical core is not purely observational, as convention enters in
  deciding which pitches belong to a scale. And a further step
  of grouping scales into named categories is a separate
  convention, which I approach through prototype theory and
  illustrate with examples from Irish music. Separating these
  layers provides a basis from which scales can be re-examined
  empirically and cross-culturally.
\end{abstract}
\vspace{0.5cm}
\noindent\small\textbf{Keywords:} musical scales | comparative musicology | ethnomusicology | 
empirical definition | prototype theory
\vspace{0.5cm}

\section*{Introduction} 

  Musical scales are near-universal in human music
  \cite{trehubHuman2000a,net00,
  brownUniversals2013,savageStatistical2015},
  and thought to be unique to it\cite{mcdmp05,fitco06,hoescheleSearching2015,
  hoescheleAnimal2017a,honingOrigins2018,phillipsMusic2025}.
  Speech and animal vocalizations are also structured sounds
  with repetition and pitch categories (e.g., calls, 
  syllables, tones), prerequisites for scales.
  But speech, even in tone languages, does not keep to a regular
  pitch over long timescales -- tones are pitched, but only between
  neighboring syllables. Even the most song-like bird
  songs do not carve up pitch space into discrete relative-pitch
  categories that can be recombined endlessly.
  Scales describe regular pitch relations that are not only
  consistent within songs, but across entire musical traditions.
  Scales are central to human music, yet there remain
  problems with how they are defined: First, disciplines disagree on
  what exactly a scale is, and terminology shifts within disciplines as
  well \cite{ellisMusical1885,barbourMusical1949,hoodEthnomusicologist1971,scholesConcise1973,nettlFolk1976,
  ekwuemeAnalysis1980a,rahnTheory1983,dowlingMusical1986,burnsIntervals1999,krumhanslCognitive2001,patelMusic2007,
  gillBiological2009,net10,tymoczkoGeometry2011,randelHarvard2012,thompsonIntervals2013,benwardMusic2021}.
  Second, different cultures can employ distinct terms, that do not translate
  cleanly across languages\cite{powersMode2001}. Third, the literature has a
  strong Western influence, which means that some definitions and assumptions
  are not suitable for all traditions. This paper attempts a reset -- a definition
  as culture-free as the question permits.

  The time is ripe for a reassessment. Automated pitch extraction
  from audio\cite{mauchPYIN2014}, source separation methods for
  isolating melody lines\cite{canoMusical2019}, and large audio
  collections covering many traditions\cite{woodGlobal2022} now make it
  feasible to compute pitch distributions, infer scales, and compare
  across cultures -- even if substantial manual supervision is still
  required\cite{mcbrideMelody2024,brownMusical2025a}.
  These tools demand an updated definition, because the
  assumptions buried in older verbal definitions start to surface
  when one tries to automate the process of scale inference. Choices must be
  made about the analysis window, about what counts as a note, about
  what counts as in-scale. The paper aims to make these choices explicit.

  Three questions help identify where definitions diverge:
  The first is \textit{what does a scale include?} A scale is
  often described as an ordered set of pitches and nothing more. But in
  many traditions, analogous terms bundle pitch together with
  tonal schema, ornamentation rules, affective associations, and contextual
  prescriptions\cite{powersMode2001}. The choice of which features to include is itself
  a property of the tradition. The same elements can in principle be
  mixed and matched, and a feature that is inseparable from pitch in one
  tradition may be entirely independent of it in another. Whether these
  richer specifications belong to the definition of a scale, or to
  something else such as a mode, a tradition, or a performance practice,
  is itself a question to resolve. 

  The second question is \textit{what does the scale refer to?} It may be
  a prescriptive abstraction, a set of pitches that a composition is
  supposed to draw from. Such an abstraction might specify
  precise interval ratios, or it might give only an arrangement of
  approximate step sizes (big, big, small, and so on). Another answer is
  that a scale is descriptive, a summary of which pitches were in fact
  used. The descriptive view itself comes in two forms. A scale can be
  a statistical regularity across a performance or collection of
  performances. Alternatively, it can be the social convention that a
  performer was aiming at. Both forms are empirical, but the latter is
  an approach that tempers observation with strong priors about
  what to expect -- formally, a Bayesian approach. The two need not
  strongly coincide. A performer can sing out of tune in the major scale,
  in which case the convention and the statistics diverge. 

  The third question is \textit{who is defining the scale?} An insider
  working from the categories of their own tradition will describe
  something different from an outside analyst extracting structure
  from a recording, or an ethnomusicologist situated between the two.
  An \textit{emic} analysis uses a tradition's own conventions, and is necessary when
  the research question depends on them -- when, for instance,
  certain scale degrees carry distinct functional roles in a ritual
  or genre. This approach is also useful when the data lead to an ambiguous
  picture of a scale: few or low quality recordings,
  ambiguous tonality, or imprecise performance can each be partially
  compensated by prior knowledge of the tradition.
  An \textit{etic} analysis, by contrast, applies
  a uniform procedure across traditions without committing to any
  one set of native conventions. It is to a degree necessary for comparative
  work, and it is the natural option where there is no
  verbalised theory of pitch, or where the performers of historic
  recordings cannot be consulted. The two stances are
  complementary: each can answer questions the other cannot.

  To answer each question in turn: 
  \textit{What does a scale include?} Here, a scale is a
  set of pitches together with a tonic, where the tonic is a
  zero-position -- the reference relative to which the other
  intervals are specified. It does not carry
  fixed melodic or tonal connotations that the term sometimes brings;
  the tonic can certainly have these relations, but they are not considered here
  to be part of the definition. This answer is the most etic part of the proposal,
  as it tries to find a common core across traditions,
  leaving behind different factors like melodic rules and ornamentation. 
  \textit{What does the scale refer to?} Being minimal, this definition
  is compatible with multiple types of scales: it accommodates scales extracted from
  instrument tunings, from single recorded performances, from song
  repertoires, or from whole traditions. Which object one recovers
  depends on the data used, and this work situates the empirical core
  in relation to other types of scales that are not directly measurable
  -- perceptual, social, mathematical.
  \textit{Who is defining the scale?} The procedure can be
  deployed both emically and etically, depending on how decisions
  are made on which pitches count as part of a scale.
  In its default form it is more etic, since the output is stripped
  to the minimum and the procedure is uniform across traditions.

  To restate the proposal: a scale, in the empirical sense used here,
  is a statistical regularity in the observed organization of pitch -- a set
  of intervals relative to a tonic, and nothing more. This minimal
  core is not a claim that pitch is all a scale is; it is a claim
  about which part of a scale takes the same form across traditions,
  and is therefore the part that can be measured and compared. It
  sits alongside other senses of the term, such as theoretical
  abstractions, or socially accepted categories.
  Convention is an important component, both in deciding which pitches
  count as part of a scale and which fall outside it, and in how
  scales are grouped and named. Separating the empirical core from
  these other layers will facilitate a firmer basis from which one
  can study how scales are, and have been, used throughout the world.

  \begin{figure*}[th!]
  \centering
  \includegraphics[width=\textwidth]{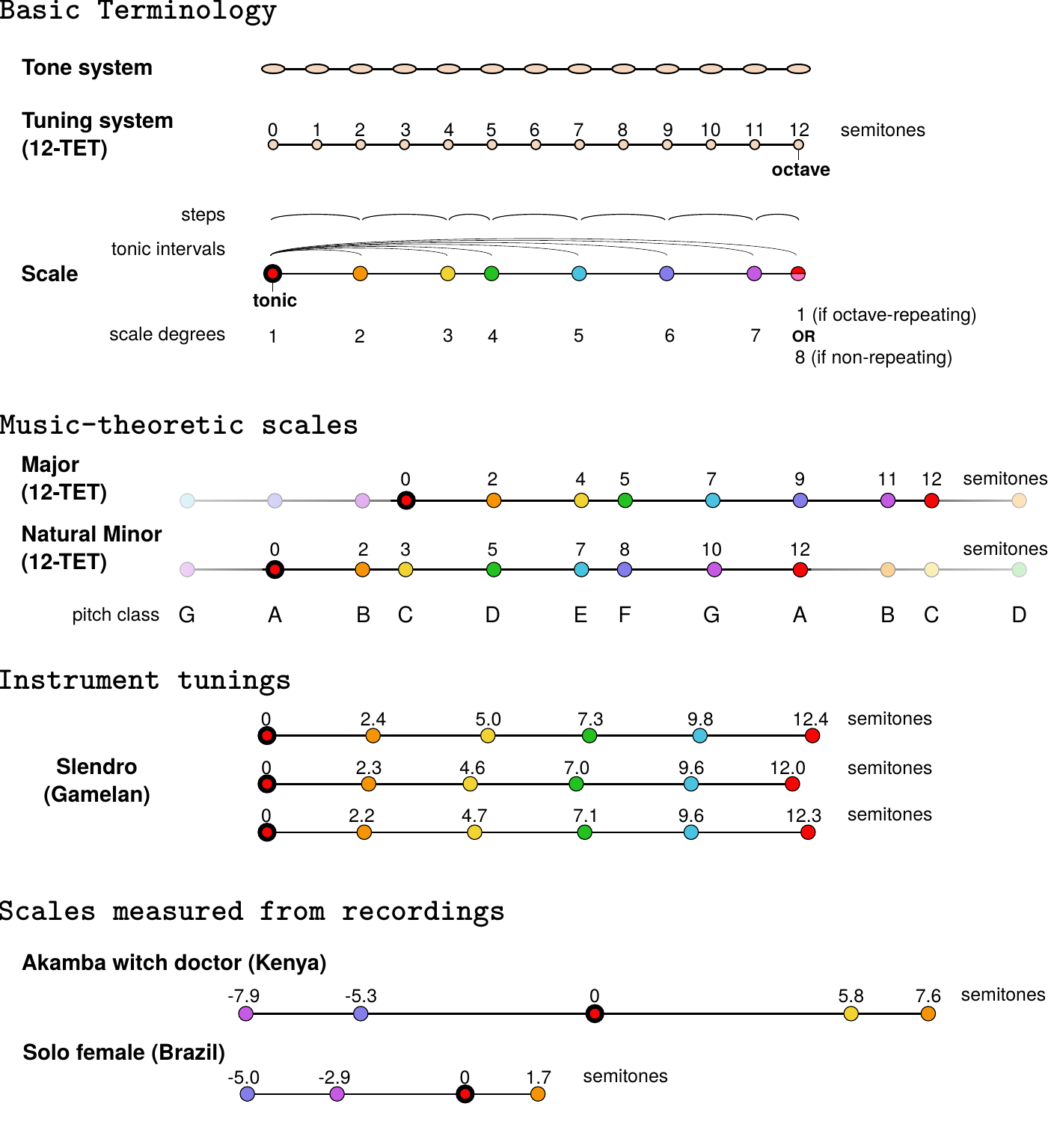}
  \caption{\textbf{An illustration of several types of scales and relevant terminology.}
  \textbf{Basic Terminology:}
  A \textit{tone system} is the set of pitch classes available within
  a tradition or on an instrument; a tone system does not specify exact intervals.
  A \textit{tuning system} assigns specific intervals to those pitches;
  the example shown is 12 tone equal temperament (12-TET), which divides
  the octave into twelve equal semitones (100 cents).
  A \textit{scale} is a tonally-anchored subset of pitches;
  a generic seven-degree scale is shown, with the tonic
  (red point with bold border) at degree 1; pitches are colored by degree.
  Arcs above the scale mark \textit{steps} (between adjacent degrees)
  and \textit{tonic intervals} (from each degree to the tonic).
  The position one octave above the tonic is degree 1 again
  for an octave-repeating scale, or degree 8 for a non-repeating scale.
  \textbf{Music-theoretic scales:}
  C major and A natural minor in 12-TET. Both belong to the diatonic
  family and share the same set of pitch classes,
  but differ in their tonic; coloured points mark in-scale pitches
  and faded points mark the same pitch classes in adjacent octaves.
  \textbf{Instrument tunings:}
  Three pentatonic slendro tunings obtained by measuring the pitches
  of individual notes on an instrument\cite{sur72}.
  These instruments have six notes but these are octave scales:
  there are only five distinct scale degrees, as the first and last notes
  are considered equivalent; the highest pitch is not exactly
  12 semitones above the first because these are measurements,
  not theoretical values. The underlying scales are octave-repeating,
  as is apparent from full-orchestra tunings (not shown).
  \textbf{Scales measured from recordings:}
  Two scales extracted empirically from sung performances\cite{brownMusical2025a},
  with intervals measured in semitones relative to the inferred tonic.
  It is not known if these are octave-repeating or not,
  or whether the pitch classes are drawn from a larger tone system
  -- one would have to study other musical examples from these traditions
  to obtain a fuller picture.
  }
  \label{fig:1}
  \end{figure*}

\section*{Glossary} 
  I start by enumerating definitions of relevant terms, which
  are illustrated visually in \fref{fig:1}.
  That means jumping ahead a bit to the proposed definition of
  a scale, which I will justify later.
  First I enumerate the standard terms:

  \begin{description}[leftmargin=16pt, style=unboxed, itemsep=0.2em,
                      parsep=0pt, topsep=0.25em]
    \item[\textit{Scale.}] A scale is a set of pitch categories,
     defined by their intervals relative to a reference pitch called the tonic.
     A scale can have any range, and need not repeat at the octave.
    \item[\textit{Pitch.}] The perceived highness or lowness of a sound,
      correlated with its fundamental frequency $f$ (in Hz). Because
      pitch perception is approximately logarithmic, it is usually more
      convenient to work with log-frequency. Log-frequency is often 
      expressed in units of cents, where $100$\,cents is one semitone in
      equal temperament and $1200$\,cents is an octave.
    \item[\textit{Interval.}] The relative frequency distance between two pitches,
      expressed as a ratio of their frequencies or, equivalently, a difference in
      log-frequency. 
    \item[\textit{Octave.}] The interval with a frequency ratio of 2:1
      ($1200$\,cents). In most traditions, pitches an octave apart are
      judged as strongly similar (\textit{octave equivalence}; see
      below).
    \item[\textit{Pitch class.}] A category of pitches treated as
      equivalent. In most Western and much non-Western music, pitch
      classes are defined by octave equivalence, so that every C
      belongs to the same pitch class.
    \item[\textit{Scale degree.}] Ordinal position within a scale,
      conventionally indexed from the tonic (1st degree) upward. 
      For non-octave scales where unique degrees can exist above and
      below the tonic, it makes sense to call the tonic the zero position.
    \item[\textit{Tonic.}] The reference pitch to which the other scale
      intervals are referred. This word is sometimes loaded with harmonic
      or melodic roles, however here I use it in its most basic sense.
    \item[\textit{Step.}] The interval between two adjacent scale
      degrees. Steps within a scale need not be equal; the major
      scale, for example, has two step sizes (tone and semitone).
    \item[\textit{Step size.}] The magnitude of a step on a log-frequency scale.
    \item[\textit{Tone system.}] A superset of pitch classes used within
      a musical tradition or available on an instrument;  also called a
      \textit{gamut}. A scale is a subset of the tone system that is
      used in a particular piece or context --
      \eg a piano has many notes that form a tone system,
      but only a specific subset -- the scale -- is used in a melody.
    \item[\textit{Tuning system.}] A specification of the exact
      frequencies (or frequency ratios) assigned to each scale degree in a scale or tone system.
      A single scale can be realised with more than one tuning system:
      the diatonic major scale can be tuned in just intonation, in
      Pythagorean tuning, or in 12-tone equal temperament (12-TET).
    \item[\textit{Equidistant scale.}] A scale whose adjacent steps
      are all of equal size in log-frequency; sometimes called an
      equal-step, equitonic, or symmetric scale. The chromatic scale in 12-TET is the
      familiar Western example; non-Western examples include the
      approximately equidistant 7-note Thai scale and the
      5-note Javanese slendro. There is no standard for how much
      a scale can deviate from the exact theoretical definition
      and still be considered equidistant.
  \end{description}

  \noindent
  Second I enumerate a set of non-standard terms, but which are needed
  to disambiguate different concepts:
  \begin{description}[leftmargin=16pt, style=unboxed, itemsep=0.2em,
                      parsep=0pt, topsep=0.25em]
    \item[\textit{Octave-repeating scale.}] A scale whose intervals are defined
      within one octave and then repeated in every other octave;
      has also been called ``octave scale''\cite{mcbrideConvergent2023b}.
      A scale with $N$ pitch classes has $N$ steps, since the closing step
      from the highest degree to the octave-equivalent of the tonic is counted.
      Most music-theoretic scales in use are octave-repeating scales, but exceptions exist
      (\eg, the Bohlen-Pierce scale, which repeats at the 3:1 tritave).
    \item[\textit{Non-repeating scale.}] A scale defined within a fixed pitch range,
      with no periodic repetition. A scale with $N$ degrees has $N-1$ steps.
    \item[\textit{Scale interval.}] An interval measured between any
      two pitches in a scale. For non-repeating scales the set of possible scale
      intervals is finite. If there are $N$ scale degrees, there are $N(N-1)/2$
      scale intervals (the number of unique intervals between all scale degrees,
      not counting unison intervals). For octave-repeating scales there are $N(N-1)$
      scale intervals smaller than an octave. One can in principle specify an
      infinite number of scale intervals across octave registers.
    \item[\textit{Tonic interval.}] An interval measured between any
      pitch in a scale and the tonic. The real-valued counterpart
      to the scale degree -- \ie, the fifth degree of a major scale
      in just intonation is equivalent to a tonic interval of 702 cents.
      This has been called `scale note'\cite{mcbrideConvergent2023b} and
      `scale pitch'\cite{scherbaumTonal2020}, 
      but I feel these names are not as descriptive as tonic interval.
    \item[\textit{Scale family.}] A set of scales that share the same
      arrangement of steps but differ in their tonic. For octave-repeating scales,
      the pitch classes have a circular structure, so the members of a
      family are related by circular permutation: changing the tonic --
      provided the steps are not all equal sizes -- yields a new scale
      with different tonic intervals. The most well-known example is
      the \textit{diatonic} family, whose members are conventionally
      called the \textit{diatonic modes}. The same logic applies to
      non-repeating scales that share a step arrangement but differ in
      tonic, even though circular permutation does not apply.
  \end{description}

  \noindent
  Third, I highlight that \textit{mode} is context-dependent and should
  be used with care. Historically the word has carried many
  meanings\cite{powersMode2001}. In one usage, modes are specific scales
  drawn from a scale family -- \eg, the diatonic modes\cite{barbourMusical1949}
  -- in which case the term is effectively synonymous with scale. In
  broader usage, a mode is a scale paired with additional musical or
  extramusical conventions: in Gregorian chant, modes assign tonal or
  melodic roles to scale degrees, such as starting and ending notes; in
  ethnomusicology, the term is often used to translate non-Western concepts that
  bundle pitch with characteristic melodic motifs, ornamentation, affect,
  or contextual associations such as season or time of
  day\cite{hoodEthnomusicologist1971}. Because of this range of meanings,
  in comparative or empirical contexts it is prudent to avoid \textit{mode}
  where \textit{scale} will do.
  
  \subsection*{Illustrative examples}

  \nsi{The major scale}
  An octave-repeating scale with seven pitch classes, with steps T-T-S-T-T-T-S.
  When paired with a specific tuning system the steps are assigned exact sizes
  -- \eg, in 12-tone equal temperament T (tone) is \SI{200}{cents}
  and S (semitone) is \SI{100}{cents}. In practice the exact intervals
  vary according to the precision of the singer or the tuning method.
  Its scale degrees are conventionally labelled
  1--7, or named \textit{do, re, mi, fa, sol, la, ti}.

  \nsi{Just intonation}
  A tuning system in which intervals are given by small-integer
  frequency ratios. Just-intonation major has scale intervals
  1:1, 9:8, 5:4, 4:3, 3:2, 5:3, 15:8, 2:1; in cents, this is
  0, 204, 386, 498, 702, 884, 1088, 1200. In this case T is not a single
  value, and can be either \SI{204} or \SI{182}{cents}.

  \nsi{Equal temperament}
  A tuning system in which the octave is divided into a fixed number
  of equal steps in log-frequency. In 12-tone equal temperament
  (12-TET) each semitone is exactly \SI{100}{cents}, giving the major
  scale degrees 0, 200, 400, 500, 700, 900, 1100, 1200.

  \nsi{Slendro}
  A pentatonic octave-repeating scale used in Javanese and Balinese gamelan music.
  \citet{perlmanUnplayed2004} describes it as a tuning system, and describes modes (in
  the sense of circular permutation, same pitches, different tonic)
  which could also make slendro a scale family -- all four definitions
  appear to be valid. The five pitches per octave are
  approximately equidistant -- steps of roughly \SI{240}{cents} --
  but the exact intervals vary considerably
  from one gamelan set to another (\fref{fig:1}, \textit{Instrument tunings}),
  with no single canonical tuning\cite{sur72}.
  Slendro illustrates two points: that scales need not derive from
  small-integer frequency ratios, and that a scale's tuning can be a
  property of a particular instrument or ensemble rather than of an
  abstract specification.

  \nsi{Scales from melodies}
  I provide two examples of a scale measured from  vocal melodies
  in \fref{fig:1}, \textit{Scales measured from recordings}.
  In these examples, I do not know of any name for the scales (someone might).
  One has a range greater than an octave, but the scale degrees
  do not repeat at octave intervals. The other has a range lower
  than an octave. These may be imprecise renditions of a more
  standard tuning system, or the scale degrees may be a subset of
  a larger tone system. Or they could be standard scales within
  each tradition. Without further recordings or ethnographic evidence all
  we have is the scale that we extract from the recording.

  \subsection*{Important concepts}

  \nsi{Transposition invariance}
  Pitch perception is largely relative: a melody played in one key or register
  is recognised as the same melody when shifted to any other. A
  scale, under the definition used here, is therefore specified by
  its intervals, not by absolute frequencies. Two performances of
  the major scale in different keys, one in C and one in G,
  instantiate the same scale.

  \nsi{Octave equivalence}
  Pitches an octave apart are treated as functionally similar in
  many musical traditions. This assumption underlies both the notion
  of pitch class and that of the octave-repeating scale. It is common
  enough that it is often taken for granted, but it is a
  perceptual and cultural regularity of listeners, not a necessary 
  part of the definition of scale.

  \nsi{Tonal hierarchy}
  Within a scale, not all pitches are functionally equivalent: some
  are heard as more stable or structurally central than others. The
  most stable pitch is the tonic; it is often, though not always,
  the most frequent pitch and the first or last note of a phrase.
  Many traditions have names for multiple notes in the hierarchy
  \cite{sig77,marcusArab1989,vijayakrishnanGrammar2007,benwardMusic2021} -- 
  \eg, the second-most-stable pitch is sometimes called the \textit{dominant}.
  In its narrowest sense the tonal hierarchy is just the relative
  prevalence of pitch classes, as observed in a pitch-class histogram
  or in listeners' probe-tone ratings\cite{krumhanslCognitive2001}.
  In a richer sense it also encompasses conventions about note
  function -- which degrees may begin or end a phrase, which are
  typically ornamented, which are unstable and require resolution.
  The two senses are related but not interchangeable, and which one
  is meant depends on the analytic frame.

\section*{Scale Definitions} 
\subsection*{Scale in the scholarly literature}
  The scholarly literature on scale is dominated by Western sources;
  the question of how the concept maps onto other traditions is
  the focus of the next subsection. Here I survey how the scale
  is defined in the literature. My original intention was to compare definitions
  across disciplines -- (Western) music theory, ethnomusicology,
  cognitive science, mathematics, and computer science --
  however this approach was unsuccessful due to the ambiguities in how scales are defined
  \cite{ellisMusical1885,barbourMusical1949,hoodEthnomusicologist1971,scholesConcise1973,nettlFolk1976,
       ekwuemeAnalysis1980a,rahnTheory1983,dowlingMusical1986,burnsIntervals1999,krumhanslCognitive2001,patelMusic2007,
       gillBiological2009,net10,tymoczkoGeometry2011,randelHarvard2012,thompsonIntervals2013,benwardMusic2021}.
  Many sources define scale only in passing, leave key
  details implicit, or omit details entirely. \citet{thompsonIntervals2013}
  goes as far as identifying multiple distinct things to which the
  word can refer (a point taken up in \textit{What does Scale refer to?}
  below). What can be said is that there is a shared core --
  an ordered set of discrete pitch categories -- and that
  authors differ on a small number of axes around it.

  \nsb{Octave inclusion.}
  Some authors explicitly include the octave in the
  definition\cite{krumhanslCognitive2001,patelMusic2007,gillBiological2009};
  only one source explicitly states that a scale is defined by a range
  that may or may not span an octave\cite{barbourMusical1949}. Most
  music-theoretic scales, Western or otherwise, assume octave
  equivalence and treat scales as bounded by an octave,
  with some exceptions where scales exceed one octave\cite{marcusArab1989,far04}.

  \nsb{Prescriptive vs descriptive.}
  Some authors treat scales as prescriptive: music-theoretic constructs
  that \textit{prescribe} which pitches are allowed in a
  composition\cite{patelMusic2007,gillBiological2009}. Others treat
  them as descriptive: summaries of the pitches actually used in a
  melody or musical style\cite{scholesConcise1973,nettlFolk1976}.
  These need not be in conflict -- the same scale can be approached
  either way -- but a definition that does not say which is meant
  is ambiguous in practice.

  \nsb{Direction.}
  Scales are most often presented in ascending direction, although most
  definitions do not specify this. In some cases scales have been ordered
  from high to low pitch\cite{kubyt85,marcusArab1989}. 
  In some cases ascending and descending forms differ\cite{randelHarvard2012,benwardMusic2021}.
  One such case in Western music is the melodic minor scale;
  there are other examples from other traditions\cite{marcusArab1989,far04}.

  \nsb{Tonic.}
  The tonic is frequently referred to in definitions but its
  necessity is rarely explicitly asserted. Authors who include it tend to do so
  implicitly, by referring scale intervals to a reference pitch;
  others omit it. I take the tonic to be part of the definition,
  for reasons later argued in \textit{Resolutions}: without it, scales
  that share a step arrangement but differ tonally -- the most
  familiar case being the members of the diatonic family -- would
  collapse into a single object, contrary to how they function in
  practice.

\subsection*{Scale across musical traditions}
  The literature surveyed above is mainly Western. The question
  for this subsection is whether the resulting definition can be applied
  beyond that literature without distortion. I argue that it can, for
  a specific reason: the definition is deliberately minimal, and a minimal
  definition can be slotted into traditions whose own concepts of pitch
  organisation are more or less detailed. The aim here is not an
  authoritative ethnomusicological survey; it is a check on compatibility,
  working through a small number of cases organised by how much explicit
  verbal theory a tradition has built up around its own pitch organisation. 

  \nsb{Minimal verbalised theory of pitch.}
  Some traditions have no native vocabulary for scale, interval, or
  pitch class\cite{ricyi80,kubyt85}, although every tradition I have read
  about uses some form of pitch-height language --
  low/high, fat/thin\cite{ricyi80}, big/small\cite{kubyt85}. The absence
  of a verbal theory does not entail the absence of a scale as a regularity
  in practice\cite{ekwuemeConcepts1974,baiyt88,slobinFolk2011,aromCategorization2019}:
  contrary to \citet{nettlMusic1977}'s claim that ``a scale does not exist
  in the mind of the native musicians'', the existence of statistical
  regularity in melodic pitch is itself evidence that some such regularity
  exists in the producers' and listeners' minds, since otherwise the
  pitches could not be arranged so regularly. Where verbal theory is
  absent, a scale can be recovered by analysing
  melodies\cite{nettlMusic1977} or through cognitive
  testing\cite{jacobyUniversal2019a,cookeReport1992,voisinMusical1994}; the present work
  is developed to support the former route. The required caveat is that in these cases
  it is prudent to be open to the possibility that the music in question
  does not organise pitch into discrete categories at all.

  \nsb{Informal naming.}
  Other traditions name individual pitches, intervals, or
  characteristic motifs without assembling them into a formal scale
  theory\cite{zemet79,kubam80,vanam80,felyt81}. The 'Are'are of the Solomon Islands,
  for instance, have native terms for specific intervals
  (\textit{rapi 'au}, \textit{hoa ni 'au}, \textit{hari 'au}) and for
  the octave (\textit{aano suri}), but no separate term for a scale
  \cite{zemet79}. Here the components of the proposed definition
  are visible in native usage even though the tradition does not state
  them as a single object; the empirical scale can be assembled from
  the named elements together with what the music actually does.

  \nsb{Direct counterparts.}
  A few traditions have native terms that map closely onto a Western-style
  scale, as a set of pitches anchored to a tonic. Examples include
  \textit{g\={a}m} in Persian music\cite{far04}, \textit{thaat} in
  Hindustani music\cite{widdessRagas1995}, \textit{melakarta} in
  Karnatic music\cite{vijayakrishnanGrammar2007}, and \textit{onkai} in
  Japanese music\cite{hughesTraditional2008}. In each case the tradition
  also has a separate, more central term for melodic type
  (\textit{dastg\={a}h}, \textit{r\={a}ga}, \textit{senp\={o}}); the
  scale-set term is secondary and in some cases imported from outside
  (Persian \textit{g\={a}m} is a loan-word from French
  \textit{gamme}\cite{far04}). 

  \nsb{Composite concepts.}
  The fourth group includes traditions whose central pitch concept
  bundles much more than pitch: \eg, \textit{mak\={a}m} in Turkish
  music\cite{sig77}, \textit{maq\={a}m} in Arabic
  music\cite{marcusArab1989}, \textit{r\={a}ga} in Indian
  music\cite{widdessRagas1995}, \textit{pathet} in Java and
  Bali\cite{perlmanUnplayed2004}, \textit{senp\={o}} in Japanese
  music\cite{hughesTraditional2008}. These concepts pair pitch with
  characteristic motifs, ornamentation rules, affective associations,
  and contextual prescriptions such as season or time of
  day\cite{powersMode2001}. They are variously translated as ``scale''
  or ``mode'', but neither translation is completely accurate.

  The scale definition proposed here is compatible in all four cases.
  Where there is no verbal theory of pitch, a scale can still be recovered
  from melodic statistics. Where pitch elements are named informally,
  or when there is a developed theory of scales,
  this information can be used to provide an empirical account that aligns
  with the insiders' conceptions of which pitches are noteworthy.
  Where the native concept bundles more than pitch, the definition
  extracts only the pitch-organisational layer and is silent on the rest.
  The additional material -- ornamentation, affect, context --
  may be real and important, but it is not the same object
  as a scale, and the question of how the two relate is taken up in
  \textit{What does Scale refer to?} below and in \textit{Social
  Convention} later. Yet one must keep in mind that a scientific account
  of pitch is but one aspect of music, which is otherwise multifaceted.

\subsection*{What does Scale refer to?} 
  As I wrote above, there are multiple ways to define a scale, and that
  people can include different sets of information. Here I enumerate
  different objects that a scale can refer to, at different levels
  of abstraction, which are not mutually-exclusive.

  \nsb{Physical/instrumental:}
  A scale is the set of pitches available on an instrument -- according
  to the definitions set out earlier, this could mean a \textit{descriptive
  scale} if the pitches are used in a piece of music, or it could be a
  \textit{tone system} from which scales are drawn.
  This definition is familiar to instrument makers and
  those musicians who tune their instruments. The first cross-cultural comparison
  of such empirical measurements was by \citet{ellisMusical1885}, who called this the
  ``instrument scale''; \citet{thompsonIntervals2013} referred to this as a
  ``physical scale''.  There are a number of ways this physical scale
  may differ from other types of scales: A melody does not have to use
  all of the available pitches on an instrument. An instrument might only have
  a subset of scale degrees as it could share a melody with
  another instrument, (\ie, a hocket melody)\cite{nketiaHocketTechnique1962,ampeneAsante2020}.
  or be used for harmonic accompaniment.
  Even fixed-pitch instruments are not strictly fixed -- \eg, on
  aerophones one can play notes in between notes using alternate
  fingerings or by partially covering holes\cite{san05,shias15}.

  \nsb{Mathematical/theoretical:}
  A scale is a set of intervals defined by frequency ratios or intervals.
  This is the view of music theorists from Pythagoras through
  to modern set theory approaches. It is precise, but is disconnected
  from practice, where intonation deviates from theoretical ideals.
  There is a large body of modern mathematical scale theory that sets
  aside frequency ratios entirely, treating scales as subsets carved
  from a presupposed chromatic set\cite{harasimAxiomatic2020}.
  These theories deal with octave-repeating scales, assume octave equivalence,
  and are based primarily on Western tonal practice\cite{tymoczkoGeometry2011}.

  \nsb{Statistical/empirical:}
  A scale is a statistical regularity in the distribution of pitch,
  operationalised as summary statistics (\eg, mode, mean) of pitch
  (or pitch class) categories in histograms. Not all pitch
  histograms represent scales. Depending on what sample is used to create
  them, histograms may only capture only a part of a scale, or 
  more pitch classes than what belongs to a single scale.

  \nsb{Psychological/perceptual:}
  A scale is a mental representation -- a set of learned pitch
  categories that individual listeners use to parse and predict melodies, and performers
  use to produce melodies. Scales exist in the mind of the listener, who
  interprets what they hear based on internal models of what they have heard before.
  The scale sounded in a performed melody is a tangible
  instantiation of the mental representation -- if recorded, it becomes a fixed
  object that one can measure. Otherwise, it can only be probed indirectly
  through cognitive experiments.
  
  \begin{figure*}[th!]
  \centering
  \includegraphics[width=\textwidth]{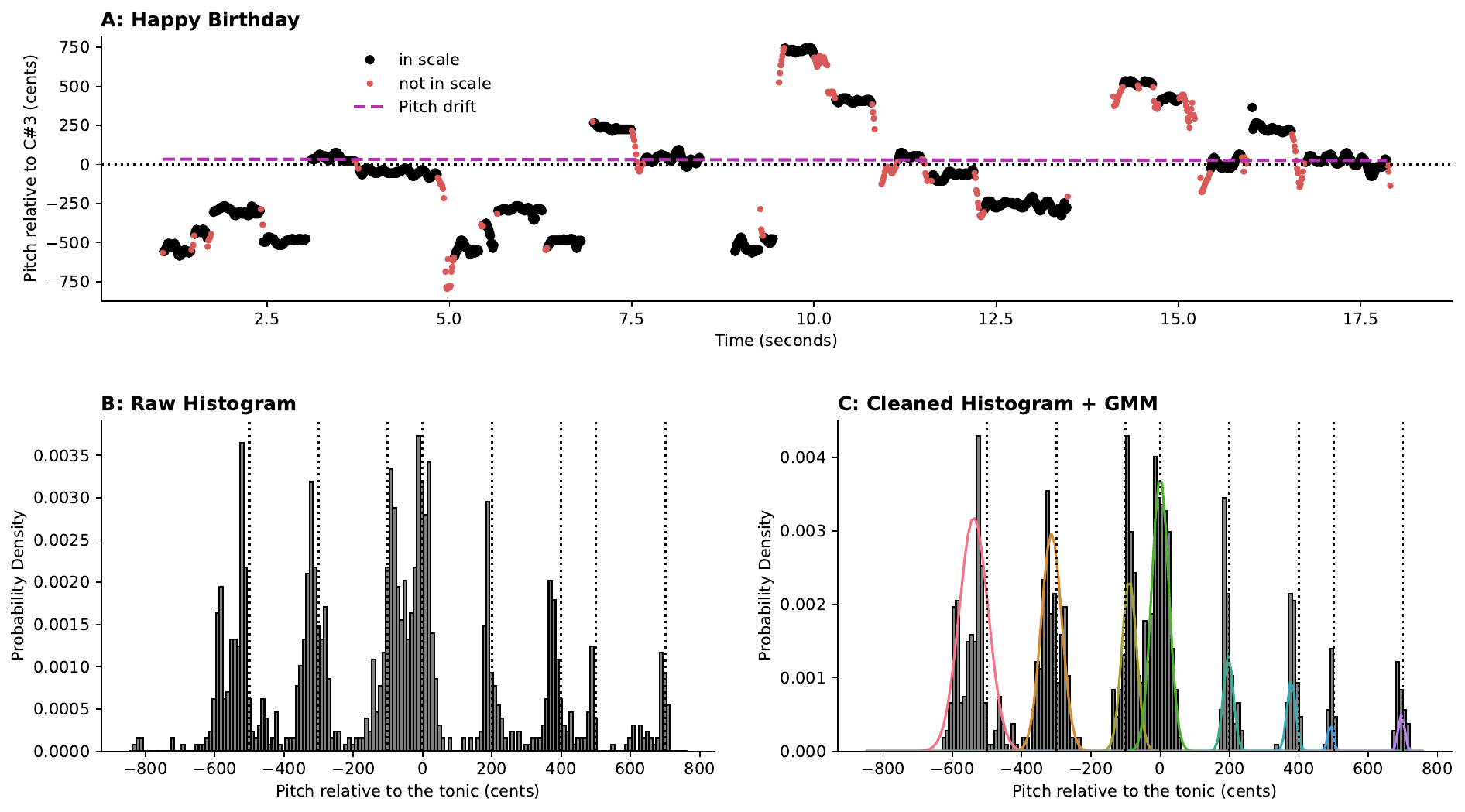}
  \caption{\textbf{An illustration of an empirical scale, using a sung rendition of ``Happy Birthday.''}
  \textbf{A:} Fundamental frequency of a sung melody over time, expressed in cents relative to C\#3.
  Pitches are assigned scale membership on the basis of being part of
  a note (black) or instead being part of a passing tone (red).
  A linear pitch-drift model with a shared slope and per-note intercepts
  is fitted to the in-scale points (dashed magenta line);
  in this case the drift is small (\SI{-0.5}{cents/s}), and is included
  as an illustration of how one can correct for drift.
  \textbf{B:} Histogram of all pitch values centred on a tonic, whose 
  precise location is estimated using a Gaussian mixture model (GMM; not shown)\cite{scherbaumTonal2020}.
  Dotted lines mark 12-TET scale degrees.
  \textbf{C:} Histogram of in-scale pitches after subtracting the fitted drift,
  with an 8-component GMM overlaid (coloured curves).
  A GMM is a natural choice for fitting scale degrees to a pitch histogram,
  as it can fit separate Gaussian curves for each pitch category, modelling
  its mean pitch, its variance, and its relative weight.
  Without assuming octave equivalence, and labelling the fourth
  component (C\#) the tonic, the tonic intervals are the means
  of the GMM components,  
  $\{ -538, -314,  -89,    0,  197,  378,  492,  696 \}$.
  }
  \label{fig:2}
  \end{figure*}

  \nsb{Social/conventional:}
  A scale is a culturally agreed-upon categorisation of pitch organisation
  -- a convention for grouping melodies and tunings. Sometimes regularities
  in pitch organisation are excluded from scales on the basis of them being
  ornamental, or rhythmic accompaniment -- regular pitch that is not considered
  part of the core melody. Sometimes the opposite occurs: Melodies often
  contain six scale degrees, and are still named after one of the seven-note
  scales. Sometimes melodies are widely associated with a scale, such
  that when a poor-pitch performer plays the melody\cite{pfordresherPoorPitch2007},
  even though the pitches do not match the scale,
  the song can still be considered to be in that scale.

  \nsb{Pedagogical:} A scale is a teaching tool -- a prescribed sequence
  of pitches that students practise to internalise a tuning system.
  In principle any scale can be used as a teaching scale.
  There may also be cases where a scale is only pedagogical -- the melodic minor fits
  into this category. It has different pitch classes depending on whether a melody
  is ascending or descending, and this is not known to be representative of melodies.
  This was investigated in works of Johann Sebastian Bach, where it was found that the choice of pitches
  depended on harmonic context rather than melodic motion\cite{towndrowMelodic1977}.
  Likewise the scale has been criticised for giving students the wrong impression about
  how it should be applied in music\cite{telescoRethinking2001}.

  \nsb{Practice-emergent:}
  A scale is what one obtains by aggregating over a tradition's
  melodies and isolating the pitch classes that are most commonly used.
  In this reading, the empirically primary objects are
  melodies and clusterings of melodies. The scale is a derived
  statistic that summarises which pitches the melodies use,
  and is only one feature of a cluster of melodies\cite{powersMode2001}.
  This fits one definition of mode, and analogous terms in other
  traditions. The relation between melodies, melody clusterings,
  and scales is taken up in \textit{Scales as Prototypes}.

\section*{An Empirical Definition of a Scale} 
  The definition proposed here is deliberately minimal:
  \begin{quote}
  a scale is a statistical regularity in the organisation of
  pitch, specified by a set of pitches relative to a tonic.
  \end{quote}
  What this means, in practice, is the following.
  Given a body of music -- a performance, a song, a repertoire --
  one extracts the fundamental frequencies of the pitches,
  converts them to log-frequency (cents), and examines the resulting
  distribution. If the pitches cluster around a small number of
  well-separated categories, then those categories constitute the
  scale. The scale is defined by the intervals between the cluster
  centres, and the cluster center of a reference pitch (the tonic).
  \fref{fig:2} illustrates this procedure on a sung rendition of
  ``Happy Birthday'': the pitch trace (A) is reduced to a histogram
  centred on the tonic (B), and a Gaussian mixture model fitted to the
  cleaned distribution recovers the cluster centres (C).
  I will refer to the excerpt of music used to construct such a distribution as the
  \textit{analysis window}: it may be a few bars, a whole song,
  or an entire repertoire. I will show below that the choice matters.
  Note that the same procedure, when
  applied to an instrument's notes, might yield a tuning system (which
  is sometimes the same as the scale).
  In the following, however, I will refer to scales.

  This definition has several features that are useful for
  cross-cultural work. It is \textit{descriptive}: it says what
  pitches were used, not what pitches should have been used.
  It is \textit{empirical}: it operates on observable, measurable
  signals, not on theoretical constructs. It does not presuppose
  octave equivalence, a specific number of notes, or any particular
  interval structure. And if one does have style-specific knowledge
  that they want to incorporate, one can easily specify prior distributions
  or pitch inclusion criteria.

  Stated this way, the definition is still underspecified.
  Extracting a scale from a pitch distribution requires a series
  of decisions about what sounds to include, how much music to
  analyse, how to handle imprecision, and where to draw the
  boundary between one scale and the next. The next two subsections
  unpack these decisions: the first lays out six problems that
  arise when the definition is applied to real music, and the
  second suggests how each may be resolved.

\subsection*{Problems with the Empirical Definition} 
 
  \nsb{1. The analysis window is too short.}
  A scale is not the same thing as a sample of pitches. A few
  seconds of a melody may visit only a subset of the full scale.
  In tonal hierarchies, a few pitches typically account for most of
  the notes in a melody and the rest appear less often or rarely,
  so a short window will often miss the rarer
  pitch classes. Identifying the scale therefore requires a long
  enough window that all of the relevant pitch classes are
  included in the sample. In practice, this typically means at
  least one complete melodic idea, and possibly more. The
  Happy Birthday rendition in \fref{fig:2}A is a useful reference
  point: out of the \SI{18}{seconds} shown, the last two phrases
  (\SIrange{8}{18}{seconds}) are needed to obtain all eight pitch classes.
 
  \nsb{2. The analysis window is too long.}
  While too short a window gives an incomplete picture, too long
  a window risks producing a composite of multiple scales.
  Boundaries exist that need to be respected: two melodies using
  different scales can be played consecutively, and tunes in some
  traditions involve modulations to different scales or
  transpositions to different keys. Detecting
  these boundaries is therefore essential.

  \nsb{3. Imprecision in singing or tuning.}
  Human voices are imprecise pitch-generating instruments\cite{brownVocal2023}.
  Likewise, instruments go out of tune, and without a standardised
  reference humans rely on memory or imprecise perceptual
  strategies to tune them\cite{mcbrideConvergent2023b}. Imprecision limits our ability to
  separate scale degrees, especially when steps are small.
  Highly-trained singers can sing intervals with standard
  deviations of \SIrange{20}{40}{cents}, while ``poor-pitch
  singers'' are much less precise (\SIrange{30}{320}{cents})\cite{pfordresherPoorPitch2007,mcbrideMelody2024}.
  For high imprecision it can be impossible to recover the intended
  scale without prior knowledge. Even with low imprecision there is
  a limit to how small a step can be reliably inferred from a
  single performance, due to overlap in the distributions of
  adjacent pitch classes.

  A related source of imprecision, specific to a cappella
  singing, is \textit{pitch drift}: the singer's or group's reference shifts
  gradually over the course of a performance\cite{mauja14,scherbaumTonal2020}.
  Drift smears the pitch distribution similarly to within-note imprecision,
  but its origin is different -- the reference itself is moving.

  \nsb{4. A scale can be defined at multiple levels.}
  Even setting imprecision aside, a single performance may not
  fully describe a scale. A performer may be aiming at a
  well-defined set of pitches understood within their tradition,
  but the statistical regularity of one performance is only one
  window onto that target. Aggregating across performances of the
  same song, or across performers within a tradition, yields
  different objects: ``the scale of this performance'', ``the
  scale of this song'', and ``the scale of this tradition'' are
  not exactly the same thing, even when they are nested. The definition
  therefore needs to be explicit about which level is meant, since
  each level addresses something different.

  \nsb{5. Not every sound in a recording is a harmonic/melodic pitch.}
  Recordings typically contain pitches that are not part of
  the melody. Sung performances include breaths and transitional
  pitches arising from continuous phonation -- intervals traversed
  while moving between sustained notes, rather than aimed-at notes in their
  own right. Spoken passages, percussion, audience sound, and
  accompaniment bleeding into a vocal mic add further
  non-melodic material. These events register in a pitch
  tracker but do not belong to the melody.
  Left in the histogram, transitional pitches broaden the peaks
  of nearby scale degrees, and sustained non-melodic sound can
  seed spurious peaks of its own. 

  \nsb{6. Ornamentation blurs the boundary of a ``note''.}
  Many musical traditions distinguish, in some form, between
  the structural pitches of a melody and the ornamental sounds
  that decorate them\cite{ibsenalfaruqiOrnamentation1978,krumhanslAcquisition1982,aromAfrican1991,
  lerdahlGenerative1983,williamsTraditional2004,saadahReframing2024,kinisDecoration2026}.
  The dichotomy is common but not
  universal: in Karnatic music, \textit{gamaka} (ornamentation)
  is described as constitutive of the melody rather
  than as a layer added to it\cite{sambamoorthySouth1964,schachterStructural2015}.
  The term ornamentation, moreover, has two distinct meanings that are
  easily conflated: a class of named, instrument-specific
  techniques (\eg, vibrato, glides, trills), and a
  functional claim that some pitches are decoration of a
  structural skeleton and could be deleted without affecting 
  its identity. A technique can serve a structural role in one
  tradition and a decorative role in another, and conflating
  the two may lead to disagreement;
  the consequences for the present definition are taken up in
  the \textit{Discussion}.
  Even where the dichotomy holds, ornamentation does not behave the same way
  in a pitch histogram across the different forms it can take on, and only
  some forms create a genuine problem.

  Ornaments that use existing scale degrees (\eg, mordents and
  trills on fixed-pitch instruments) are
  compatible with the scale by construction. Ornaments
  that do not linger anywhere in pitch space -- brief grace
  notes, rapid glides  -- broaden the peaks of nearby scale
  degrees but do not produce salient peaks of their own.
  The remaining case is the difficult one.
  Decorative events that pause on a pitch or that recur systematically
  can produce histogram peaks difficult to discriminate from
  scale degrees on purely statistical grounds: a sustained
  ululation, a prolonged portamento, a consistently used
  passing tone, an out-of-scale note used for emphasis (an
  ``accidental''), or vibrato wide enough that a single
  intended pitch shows up as two separate peaks.

  The distinction the analyst needs to make in this third case
  cannot be directly inferred from the pitch histogram. It is 
  \textit{functional}: which pitches form a melodic backbone that a
  listener attends to, and which serve a different role --
  textural, rhythmic, expressive, or, as in gamaka,
  simultaneously ornamental and structural. The empirical
  definition therefore needs more information than just the aggregate
  distribution to handle these cases.

\subsection*{Resolutions} 
 
  \nsb{1. A hierarchical representation of a scale.}
  A scale cannot be fully separated from its tonal hierarchy. The tonic
  anchors the scale and gives each degree its functional identity;
  without it, circular permutations of the same step pattern are
  indistinguishable. The clearest illustration is the diatonic family:
  C major and A natural minor share the same seven pitch classes but
  are not the same scale, because their tonics differ. Modulating between the two
  leads to melodies with different pitch distributions, different intervals
  of stability, and different notes at phrase boundaries. The same
  point holds for non-repeating scales, where there is no circular
  permutation to invoke: two scales with the same step arrangement
  but different tonics still differ, because the tonic determines
  which intervals are heard as structurally prominent and where the
  melody gravitates. The tonic must therefore be part of the definition
  of a scale, not an optional annotation.

  The unequal distribution of pitches in tonal hierarchies, noted above, has a
  useful consequence for the window problem. A short excerpt does
  not completely fail to capture the scale -- it captures the
  \textit{top} of the hierarchy, the most frequent pitch classes,
  missing the rarer degrees. One can take the position that an
  incomplete scale is not a scale, or one can accept the
  hierarchical nature of a scale and treat the more common subset
  as an object worth studying in its own right.

  This hierarchical structure also clarifies the status of
  occasional out-of-scale pitches. In Western tonal music,
  brief chromatic deviations are common but are often not treated
  as evidence for a different scale or a modulation. In the absence of
  an explicit change in harmonic accompaniment, this may be because
  they sit below an implicit threshold of salience.
  This may be that they are too rare
  to register as peaks in a pitch histogram constructed from
  a large corpus -- \ie, what matters may be how often the pitches occur
  across songs. The threshold is a convention within a
  tradition, but it aligns with the statistical reality --
  structural tones dominate, and are considered part of a scale,
  while other tones are rare out-of-scale deviations.

  \nsb{2. Musical structure signals scale boundaries.}
  Modulations and transitions between songs are signalled by musical structure:
  \eg, periods of silence, changes in rhythm, new melodic motifs, new harmonic context,
  phrase boundaries. They can also be identified from the melody alone:
  as a listener has expectations about what they will hear next,
  a modulation can be detected as a point where those expectations are violated
  and a new mental prediction model (in the new scale or key) leads to more
  successful predictions. Thus a model of scales requires more than
  aggregate statistics in a pitch histogram -- it requires attention to the
  melody (or harmony) itself.
 
  The practical implication is that scale analysis should proceed
  in two stages. First, segment the music into structurally
  coherent analysis windows (phrases, sections, songs). Then, analyse the
  pitch distribution within each window. This is the order in which
  a listener processes music, and it should be the order in which
  an algorithm does so too.

  Another approach that has been used by theorists in a few
  traditions is to describe \textit{compound scales}, where a single
  object describes more than one set of pitches that can be switched
  within a single piece of music\cite{sig77,marcusArab1989}.
 
  \nsb{3. Recognition corrects for imprecision.}
  When a listener recognises a melody -- through its lyrics,
  its contour, its rhythm, or its phrase structure -- the
  listener can infer the intended scale even from an imprecise
  performance. A song sung badly in the major scale is still
  heard as being in the major scale, because the listener's
  prior knowledge of the melody overrides the noisy pitch
  evidence. The listener is effectively performing Bayesian
  inference: a strong prior (knowledge of the canonical melody)
  compensates for a weak likelihood (imprecise pitch production).
 
  This means that scale identification in practice is not
  purely bottom-up. It is aided by top-down knowledge of the
  piece, the tradition, and the conventions of the musical
  style. The scale is partly recovered through recognition
  of the melody it belongs to, not just from the pitches
  themselves. This is not a defect of the definition but
  a feature of how scales function in real musical
  communication: they are robust to production noise because
  listeners bring expectations to the signal. However, this
  reliance on listener expectations has an important corollary:
  when the melody is unknown and the performance is imprecise,
  the scale may become genuinely ambiguous. A set of
  poorly-tuned pitches from an unfamiliar tradition may
  not support confident scale identification by an outside
  observer, even though an insider might have no difficulty.
 
  Drift has a more mechanical solution. Because it is systematic
  rather than random, it can be estimated and removed: fit a
  linear (or higher-order) trend to the in-scale pitches over
  time, then subtract it before constructing the histogram
  \cite{scherbaumTonal2020}. An example of this is shown in \fref{fig:2}.

  \nsb{4. Aggregation levels.}
  That a scale can be defined at multiple levels is not so much a
  problem as an analysis choice. Scale analysis operates at three
  nested levels, each answering a different question:

  \begin{description}[leftmargin=16pt, style=unboxed, itemsep=0.2em,
                      parsep=0pt, topsep=0.25em]
    \item[\textit{Single performance.}] The pitch clusters of one
      person singing one song -- or a part of a song -- on one
      occasion. This is the noisiest estimate, but the most
      specific: it captures the idiosyncratic rendition of one
      performer. 
    \item[\textit{Song.}] Aggregated across multiple performances
      of the same piece, averaging out individual imprecision and
      non-structural pitches. This gives the idealised scale of
      the composition.
    \item[\textit{Tradition.}] Aggregated across many songs and
      performers within a musical style. Aggregating all songs
      gives the tuning system of a tradition, while songs separated
      into clusters that use the same pitches can reveal specific scale types.
  \end{description}

  Each level of aggregation increases statistical power but
  decreases specificity. A tradition-level analysis is precise
  but blurred: it averages over modulations, stylistic
  differences, and individual variation. A single-performance
  analysis is noisy but faithful to what was actually sung. A
  comprehensive account of scales should be explicit about which
  level is being discussed.

  \nsb{5. Isolate the melodic stream before analysing pitch.}
  Most non-melodic content of a recording can be removed with
  signal-level tools that operate before any pitch-distribution
  model is fitted. The melody source can be isolated by source
  separation or by manually selecting segments in which only
  the melodic voice is active. Within those segments,
  transitional pitches from continuous phonation can be excluded
  by manual annotation, or for some musical styles an automated
  approach that looks for stable pitch regions may be
  appropriate\cite{rosenzweigDetecting2019}. The red points in
  \fref{fig:2}A are an example: transient pitches between
  sustained notes, labelled out-of-scale by annotation.

  \nsb{6. Salience and regularity.}
  The first two cases of ornamentation identified above are
  handled by the histogram itself: ornaments that fall on
  existing scale degrees, and transient pitches that do not
  lead to new histogram peaks. The 
  third case -- ornamentation that produces its own peaks --
  requires more attention. The goal
  here is not to recover ``structural pitch'' as such, but to
  recover the pitch set whose statistical regularities
  organise the music; whether a given peak then counts as a
  scale degree, an accidental, or an ornament is a further
  judgment that the etic procedure can only partly make.
  Insider knowledge of convention is ultimately what fixes
  scale belonging from within a tradition, and the etic and
  emic analyses will not always converge.

  A useful reframing is to characterise each pitch class along
  two axes rather than asking the binary question of whether
  it is structural. \textit{Regularity} is the peakedness of its
  histogram cluster. \textit{Salience} is a composite of total
  duration, frequency of occurrence, accentuation, stability across repeats
  of a motif, and position within a phrase (cadential and
  initial pitches weighted higher than mid-phrase ones).
  A third axis enters when more than one recording
  is available: whether the pitch is salient and regular
  \textit{across} recordings or specific to one of them.

  These axes can be drawn to identify five cases, where scale-degree
  labels are assigned based on thresholds along these axes --
  these can be chosen in line with convention (emic),
  or in line with the aims of the analyst (etic).
  \begin{description}[leftmargin=16pt, style=unboxed, itemsep=0.2em,
                      parsep=0pt, topsep=0.25em]
    \item[\textit{Regular, salient, consistent across recordings.}]
      The unambiguous case: a scale degree.
    \item[\textit{Regular and salient within a recording, but not across.}]
      An accidental -- a stable pitch in this performance,
      not a scale degree of the song or tradition. Whether to
      promote such a pitch to a scale degree at all depends
      on a threshold along the cross-recording axis: how
      often, and in how many recordings, must a pitch recur
      before it counts?
    \item[\textit{Regular but low salience.}]
      A tone sounded only once per song has
      a clear cluster centre but may fall below a salience
      threshold; the etic procedure may miss it, while
      an insider does not. For example, in the song
      \textit{Happy Birthday}, the fourth scale degree (the
      last 'Happy'; at \SI{492}{cents}) is statistically
      marginal but musically essential (\fref{fig:2}C).
    \item[\textit{Salient but not regular.}]
      Ornament by judgment: sustained ululations; slides
      whose pitch dwells noticeably without ever stabilising
      at a single value; and vibrato wide enough to produce two
      histogram peaks for what is, perceptually and
      intentionally, a single pitch. What counts as
      ``regular enough'' is a threshold on the first axis.
      In practice a practitioner might make the judgment based
      on pitch function and convention, not simply pitch regularity.
    \item[\textit{Neither regular nor salient.}]
      Trivially excluded.
  \end{description}

  In my own practice, it has not been possible do determine
  the salience-regularity profile from simply reading a
  pitch histogram\cite{mcbrideMelody2024,brownMusical2025a}.
  I iterate between the time-domain signal (pitch trace, audio) and the
  histogram, deciding what counts as a note event
  before any model is fitted -- in the cited examples ambiguous cases were
  handled by three analysts. I have come to think of this as
  \textit{joint transcription and scale inference}:
  transcription not in the sense of producing Western notation
  but in the sense of identifying note onsets, offsets, and
  ornament boundaries from the signal. There is no standard
  reference for this procedure, and other analysts may reasonably
  proceed differently -- indeed, even with a standardized procedure,
  experts within a tradition can produce different
  annotations\cite{proutskovaVocalNotes2026,chibaWhat2026}.
  The point I would defend is that scale inference involves musical judgment
  at the per-event level, not only statistics at the aggregate level.

  \begin{figure*}[th!]
  \centering
  \includegraphics[width=\textwidth]{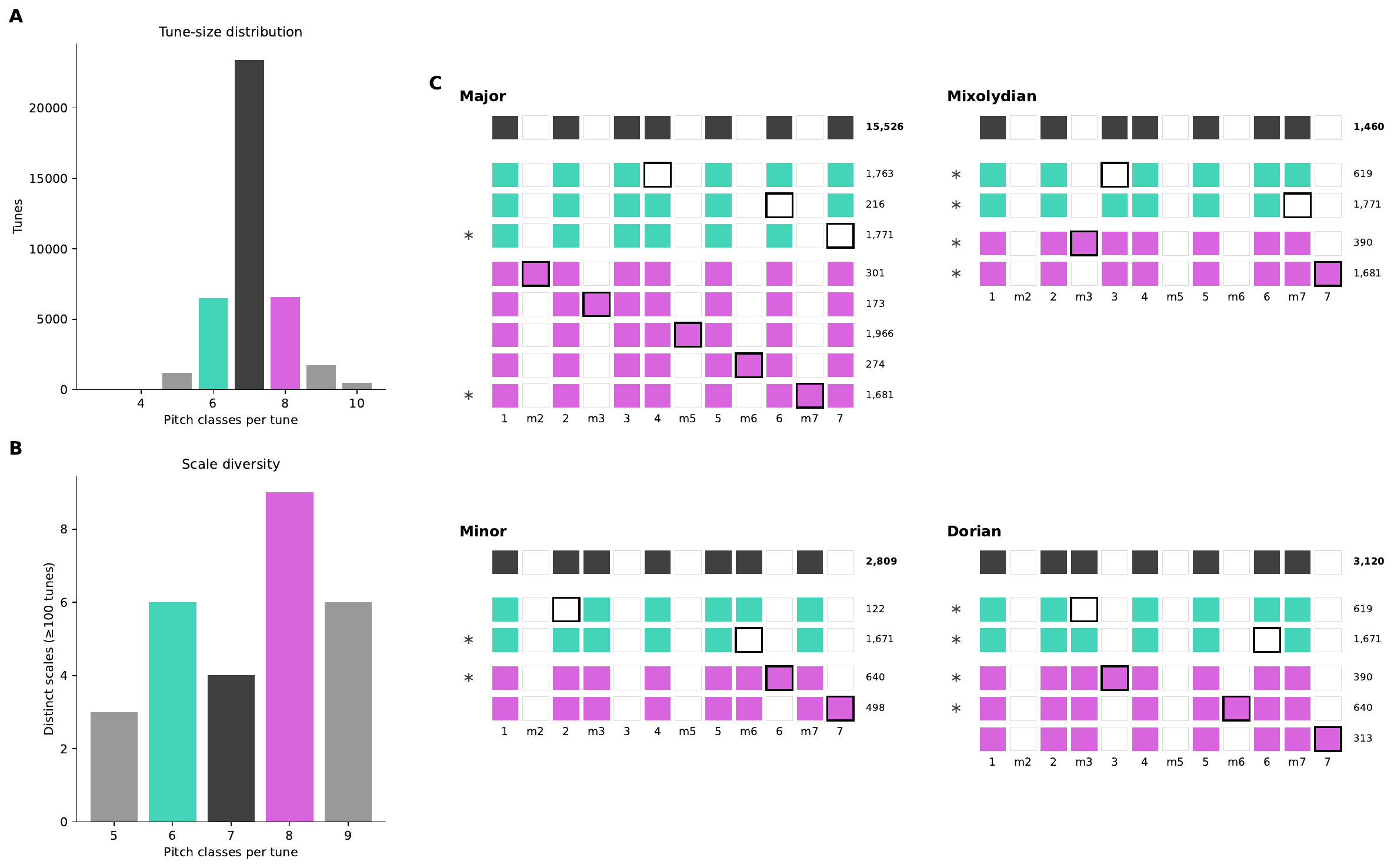}
  \caption{\textbf{Clustering of Irish melodies into four named scales.}
  \textbf{(A)} Number of pitch classes per tune in a dataset of \num{40000} Irish
  dance tunes.
  \textbf{(B)} Number of unique scales of each size that are in at least \num{100}
  melodies.
  \textbf{(C)} The 6-note (teal) and 8-note (purple) scales are all one scale degree
  from at least one of the four canonical 7-note scales (black); missing/extra scale
  degrees are highlighted by a black border. Total counts are shown to
  the right; ($^*$) some scales are similar to more than one canonical mode.
  }
  \label{fig:3}
  \end{figure*}

\section*{Scales as Prototypes}
  The empirical definition treats a scale as a statistical
  regularity in the organisation of pitch. A regularity, however, is not
  yet a named category. The following example illustrates how the
  empirical definition maps on to a category in a non-trivial way.
  Two performances can yield closely
  related pitch distributions, with the same tonic and the same
  tonal hierarchy, and still differ in whether some peripheral
  degree appears at all. Consider a tune that uses C, D,
  E, F, G, and B, but never A. The empirical procedure recovers
  a six-note scale whose pitch classes coincide, up to one
  missing degree, with C major. Whether to call this ``C major
  with a missing sixth'' or to count it as a distinct six-note
  scale is not a question the statistics decide.

  Different theories have been proposed for how humans categorize
  the objects they encounter. Of these, I consider prototype
  theory useful for describing how humans might operate when categorizing
  scales and melodies; note that I am not advocating for the specific details
  of this cognitive theory, and alternative theories of probabilistic
  categorization could also fit\cite{nosofskyAttention1986}.
  Prototype theory states that the goal of category systems is
  to ``provide maximum information with the least cognitive effort'',
  and that this is achieved when the ``categories map the perceived world 
  structure as closely as possible''\cite{roschPrinciples1978}.
  Three features of the theory are relevant here.
  \textit{Graded membership}: category members differ in
  how central or peripheral they are, with no sharp boundary. In Europe a
  robin is a more central bird than a penguin, but both are
  birds. \textit{Family resemblance}: members share overlapping bundles
  of features without all sharing any single feature;
  no single property is necessary (\eg, the ability to fly).
  \textit{Basic-level categories}: among the levels at which
  something can be categorised (\textit{recliner} /
  \textit{chair} / \textit{furniture}), one level is
  cognitively privileged -- it provides the most information for the least
  effort, and is thus the default for description.
  Throughout, \textit{prototype} refers to the central
  representative of a category, not to the category itself.

  Each of these features maps onto how named scales are used.
  Graded membership: the six-note tune above is a
  peripheral instance of a major category; a tune that uses all seven
  diatonic degrees with the expected tonal hierarchy is the
  central prototype. Family resemblance: across recordings, instances of
  the major scale share overlapping bundles of pitches, tonal
  hierarchies, and melodic conventions, but a given recording
  can fail any one of these without losing membership.
  Basic-level categories: in Western music, two named scale categories
  -- major and minor -- function as basic-level categories. Any 6- and 8-note
  variants sit at the subordinate level, treated as deviations
  rather than categories of their own.

  Within this frame, two pairs of terms become useful. A
  \textit{scale category} is a basic-level grouping of scales,
  organised around a central representative; the
  \textit{scale prototype} is that central representative --
  the most typical member of the category. Peripheral members
  differ from the prototype on rare scale degrees.
  \textit{Melody categories} and \textit{melody prototypes}
  are the analogous pair one level lower in aggregation: a
  melody category is a basic-level grouping of tunes related
  by family resemblance (\eg, in motifs, ornamentation, contour)
  and a melody prototype is its central instance.

  Two implications follow that the next section develops.
  First, the work of \textit{naming} a scale is the work of
  assigning a regularity to a category, and the assignment is
  graded rather than binary. Second, traditions may differ in
  which level they treat as basic -- some name many
  fine-grained scale categories, others a small inventory --
  and the choice of basic level is itself conventional.

\section*{Social Convention} 
  Many of the choices that go into identifying a scale -- which
  peaks count as scale degrees, where the threshold between
  structure and decoration is set, how many notes a scale is
  supposed to have, how many distinct scales a tradition recognises
  -- are ultimately based on convention. The empirical approach does
  not eliminate convention but it does let us treat music from
  different traditions in a comparable way. The same statistical
  procedure can be applied uniformly, and the conventions of each
  tradition can then be examined alongside the statistics rather
  than baked into them.

  It is useful to distinguish two points at which convention enters.
  The first is \textit{scale membership}: whether a candidate
  pitch belongs to the scale at all. The second is \textit{scale
  typology}: how the resulting scales are grouped into a small
  inventory of named types, and at what granularity.

  \subsection*{Scale membership: are pitches in or out?}
  The earlier section on ornamentation described each candidate pitch
  class along axes of regularity, salience, and cross-recording consistency.
  Convention enters in deciding where to set the thresholds along
  each axis: how peaked a histogram cluster must be to count as a
  scale degree, how often and how prominently
  a pitch must be used to count as structural rather than
  decorative, and how frequently a pitch must recur across
  performances before it is promoted from accidental to scale
  degree.

  I suspect that these thresholds are not arbitrary. They are shaped in part by
  perceptual constraints and by the statistics of a tradition's
  melodies. Like any taxonomy, decisions of where to draw boundaries
  between clusters of similar objects can be more or less optimal,
  given the cognitive resources, biases and functional needs at hand.

  A listener familiar with a tradition has internalised its tonal
  hierarchy -- a prior over which pitches are likely to serve a
  structural role. This allows the salience threshold to be lowered
  for a rare but expected pitch, as with the fourth scale degree in
  \textit{Happy Birthday}; an outsider, lacking the
  prior, might reasonably class the same note as a decorative
  accidental or a mishap.

  Pitch statistics are not the only input. Rhythmic and metric
  position, lyric emphasis, dynamic or timbral marking, and
  instrumental or pedagogical convention can all raise a pitch's
  salience above what its raw frequency of occurrence would
  suggest. Cues can even be extramusical -- a note may be
  made prominent by the word sung over it, or granted weight by
  the tradition's verbalised theory -- and they likewise shape
  where an observer sets the threshold.

  I next deal with the question of what determines the decision to
  classify a system as one scale plus or minus some notes, versus multiple scales.

  \subsection*{Scale typology: how many? which ones?}
  A second layer of convention groups the resulting scales into a small
  set of named prototypes. For example, Western theory
  groups \{C, D, E, F, G, B\} with C major rather than treating it
  as a category in its own right because it carries a seven-note
  prior: scales come in seven-note packages, and incomplete
  instantiations are matched to the nearest complete template. An
  observer without this prior -- an algorithm, or a listener from a
  tradition that does not use heptatonic scales -- would have no
  reason to posit a missing A, and might classify the melody as a
  distinct six-note scale. The judgment is not about whether A is
  in the scale but about which named category the scale belongs to.

  \nsb{Major/minor: convention finds a real binary.}
  An example of a real boundary is the split between major and minor.
  In a global corpus of vocal scales, \SI{61}{\%} contained a third
  near either \SI{300}{cents} (a minor third) or \SI{400}{cents}
  (a major third) above the tonic\cite{brownMusical2025a}.
  However, the two rarely co-occur -- only \SI{9}{\%} of scales
  included both scale degrees. If we run the same calculation for
  major thirds and fourths (\SI{500}{cents}), we find double the
  rate (\SI{20}{\%}) of co-occurrence.
  Overall, the major/minor split is the most salient.
  The third degree therefore functions to a degree as a pivot, splitting
  many of the world's scales into two broad families. 
  The decision to label one family ``major-type'' and the other ``minor-type'' is a
  convention imported from Western theory; an observer from a
  different tradition might draw the boundary elsewhere, or not at
  all. But the \textit{existence} of the boundary is not arbitrary.

  \nsb{Irish modes: convention sets the prototype count.}
  In a set of \num{40000} Irish dance tunes\cite{keithTheSessionData2024},
  \SI{58}{\%} use 7 pitch classes, while sets of 6 or 8 are each found in
  \SI{16}{\%} of tunes (\fref{fig:3}A).
  Despite the dominance of tunes with 7 pitch classes, there are fewer distinct 7-note scales
  than 6-note or 8-note scales (\fref{fig:3}B), after thresholding
  to remove rare scales. Each of
  the 6- and 8-note scales differs by only one degree from one (or more) of the four
  recognised modes: major, natural minor, mixolydian, and dorian (\fref{fig:3}C).
  In prototype-theoretic terms, the four modes describe scale
  categories, each with a central seven-note prototype and a
  periphery of 6- and 8-note variants that differ from the
  prototype on a single degree. The four-mode partition is
  then a basic-level inventory: it covers what musicians play
  with a small label vocabulary, treating the 6- and 8-note
  variants as peripheral members of the modes rather than as
  categories of their own. A subordinate-level inventory -- the
  \num{19} or so distinct scales that survive thresholding --
  has been used academically\cite{fleischmannSources1998},
  but it is not the level at which the tradition labels its tunes.

  \nsb{Granularity: basic-level categorisation as a conventional choice.}
  The Irish four-mode partition is one resolution of a more general
  choice: at what level should the named inventory carve up
  practice? Two scales differing on a single rare degree can be
  treated as separate categories or as central and peripheral
  members of one. Transpositions of the same intervallic structure
  can be counted as one category or as many. Scales are defined today
  as transposition-invariant, but historically this was not
  always the case. For example, early Arabic theory
  enumerated several hundred named scales, in part by treating
  transpositions as distinct categories; later theorists and
  pedagogues, on grounds of simplicity and teachability, collapsed
  transpositions together and absorbed many variants as accidentals
  on a smaller core\cite{marcusArab1989}. Similar stories unfolded
  in Indian and Turkish music theory\cite{sig77,widdessRagas1995}. The
  empirical statistics of the underlying melodies need not have
  changed for the named partition to shrink dramatically. The level
  at which a tradition treats its scales as basic is itself a
  conventional choice, and it can be revised on grounds that may have
  little to do with how well the partition fits the music.

  \nsb{Why 5 and 7, but not 6?}
  Across traditions with verbalised music theory, and across the design
  of fixed-pitch instruments, 6-note scales are systematically
  disfavoured relative to 5- and 7-note ones\cite{mcbrideConvergent2023b}.
  This has so far evaded explanations. One explanation is that it may
  simply be an optimal solution to scale type classification, given
  statistical tendencies for melodies to have a certain number of
  scale degrees\cite{mcbrideInformation2024}. Another hypothesis
  is structural: equidistant 5- and 7-note scales
  contain an interval close to a perfect fifth, while equidistant
  6-note scales do not. Equidistant scales have a single step size which
  may make them easier to reliably tune -- they thus may have served
  as historical precursors, in which case the presence or absence of a
  fifth could have selected for 5 and 7 over 6\cite{mcbrideCrosscultural2020,mcbrideMelody2024}.
  A more speculative alternative is that 5 and 7 are favoured simply
  because they are prime, and humans have an attraction to small primes
  in cognitive categorisation\cite{loconsoleAre2022}.

  \subsection*{Convention need not be currently optimal}
  The Irish four-mode partition and the cross-cultural 5/7 pattern
  suggest that conventions efficiently summarise practice.
  This may not always be true. A partition optimal at one time can
  ossify as the tradition shifts around it, and historical
  contingency -- pedagogy, theoretical taste, instrument design --
  can sustain a partition because it is taught and built into the
  tools, not because it is an optimal description of current melodies.

  Histories of Indian, Arabic, and Turkish scale theory\cite{sig77,marcusArab1989,widdessRagas1995}
  document several mechanisms by which named partitions change:
  invention of new modes by theorists or composers; contact with
  neighbouring traditions; disuse, with scales persisting in
  theory after dropping out of practice or vice versa; and
  theorist-driven simplification on grounds of pedagogical
  economy rather than fit to current melodies, as in the Arabic
  example above.

  Statistics alone therefore cannot tell us whether the Irish
  partition shapes current practice, was shaped by it, or both
  reflect independent pedagogical and instrumental constraints;
  the 5/7 pattern is similarly underdetermined between a cognitive
  attractor and convergent theorist simplification. Its breadth
  across traditions with limited historical contact still favours
  a cognitive reading, but the safer inference is that conventions,
  like the statistics they partition, are shaped by a mixture of
  perceptual, practical, and historical pressures that the
  available data cannot fully separate.

\section*{Discussion} 
 
  \subsection*{Summary of the definition}
  The definition has two parts, one empirical and one conventional.
  The empirical core is a statistical regularity in melodic pitch,
  expressed as a set of intervals relative to a tonic.
  It is observer-independent, operationalisable
  through pitch distributions, and -- being deliberately minimal --
  portable across traditions.

  The conventional layer covers two separate decisions, and they are
  not of equal weight. The consequential one is which pitches count
  as part of a scale and which are treated as decoration, accident,
  or noise; this shapes the empirical output directly, and is where
  insider knowledge of a tradition has the most impact. The second is
  the grouping of scales into named categories. Names matter for
  communication, but naming is a lossy summary of practice: the same
  name often spans distributions that differ measurably.
  We can cluster scales into named categories in a separate process after the
  correct pitch material has been used to infer the empirical scale.

  The empirical core can apply to multiple objects at different 
  levels of aggregation -- the scale of an instrument, a performance,
  a song, or a tradition. It also sits alongside, rather than
  replacing, the other senses in which the term is used:
  instrumental tunings, perceptual categories, and
  socially named pitch sets. Being explicit about which object is
  meant, and at which level, is what allows disagreements about
  scales to be located rather than talked past.
 
  \subsection*{Tonic: necessity and ambiguity}
  The definition adopted here includes the tonic, for reasons
  argued in \textit{Resolutions}: C major and A natural minor are
  different scales because they have different tonal hierarchies,
  even though they share a pitch set. When we want to refer to a
  pitch set without committing to a tonic, the right term is
  \textit{scale family} (see Glossary). This is consistent with
  how musicians already speak: ``the diatonic scale'' refers to
  the family, while ``the major scale'' refers to a specific member.

  For the empirical analyst, the tonic must be inferred from
  the data. It is often the most frequent pitch, or the pitch
  that begins and ends phrases. However these are not rules that
  are always followed -- until there is a more robust definition,
  one is left to decide based on an intuitive sense of salience.
  In cases where the data do not support a confident assignment,
  scales can be analysed at the level of scale intervals, as these
  are the same for all scales in a scale family\cite{mcbrideMelody2024}.

  \subsection*{Modes, melodies, and aggregation}
  The more broad sense of mode -- a scale together with
  characteristic motifs, ornamentation rules, affective
  associations, and contextual prescriptions -- can be
  viewed through prototype theory as a \textit{melody category}.
  A category organised around central tunes that display a set
  of characteristic features, with peripheral members
  that more or less resemble the prototype. This is a basic-level
  category defined at the level of melody rather than scale.
  For example, two \textit{r\={a}gas} built on the same \textit{th\={a}t}
  are distinct melody categories whose pitch marginals happen to
  coincide. 

  This bears on a long-standing intuition, expressed across
  traditions and across analytic schools, that the practical
  units of musical thought are melodies rather than abstract
  pitch sets\cite{sig77,marcusArab1989,widdessRagas1995,powersMode2001}.
  The prototype frame makes
  the intuition precise: melody categories are typically
  basic-level for performers and listeners, and scale
  categories are derived by marginalising over them.
  The primary consequence is that when comparing scales of
  different traditions, we should compare them at the same
  level. A conventional music-theoretic scale from one tradition
  cannot be directly compared with the scale inferred from
  a single performance from another tradition -- one
  is a scale derived from a melodic prototype, while the
  other is a representation of a melody that may be close
  to or far from its melodic prototype. Both objects are worthy
  of study, but they should not be conflated.

  \subsection*{Scales and overlapping concepts}
  There are distinct terms that sometimes refer to the same thing
  as the definition proposed here for scales:
 
  \begin{description}[leftmargin=16pt, style=unboxed, itemsep=0.2em,
                      parsep=0pt, topsep=0.25em]
    \item[\textit{A tuning system}] specifies the
      exact frequencies assigned to scale degrees. The same scale
      (major) can be realised in different tuning systems (just
      intonation, equal temperament). The scale is the pattern of
      intervals; the tuning system is the precise specification.
      A scale and a tuning system can be the same thing if a scale
      uses all of the available pitches in the tuning system.
    \item[\textit{A tone system}] is the superset of
      pitches available within a tradition or on an instrument. A
      scale is a subset of the tone system that is used in a particular
      piece or context.
      A scale and a tone system can be the same thing if a scale
      uses all of the available pitches in the tone system.
    \item[\textit{A mode}] is a word with many possible meanings, of which one
      is exactly equivalent to scale -- the scales of a scale
      family such as the diatonic modes.
    \item[\textit{A key}] commonly specifies both a scale and a tonic
      pitch in absolute frequency space (\eg, C major). The scale is
      invariant under transposition; the key is not.
    \item[\textit{A scale category,}] as
      developed in \textit{Scales as Prototypes}, is a basic-level
      grouping of scales: a region of scale space anchored by a
      central representative (the \textit{scale prototype}), with
      peripheral members differing on rare degrees. A scale
      extracted from a single performance is one member of a
      scale category, but the category itself is a grouping of
      scales, not a single scale. The scale prototype, by contrast,
      is a single scale -- the most typical member of the category.
  \end{description}

  \subsection*{Structure and ornament}
  The word ``ornamentation'' covers two distinct things in the
  music literature, and they are easily conflated. The first describes
  a class of named vocal and instrumental techniques
  -- Irish \textit{rolls} and \textit{cuts}\cite{cowderyMelodic1990},
  Karnatic \textit{gamaka}\cite{pearsonCoarticulation2016},
  Bulgarian \textit{tresene}\cite{ricyi80},
  guqin portamento\cite{henbingGesturebased2007}, vibrato, slides --
  recognisable devices that traditions include as part of 
  their styles. The second meaning is functional: a claim that some
  pitches are decoration of a structural skeleton and could be
  deleted without disturbing its identity\cite{osuilleabhainCreative1990}.
  The two meanings often co-occur, but they are independent. A named
  technique can carry the identity of the music in one tradition and
  serve a decorative role in another\cite{henbingGesturebased2007,
  pearsonCoarticulation2016}.

  The functional sense, with its assumed structural-ornamental
  binary, is supported by studies on melodic evolution, where
  non-accentuated pitches or grace notes tend to change often
  across repeated performances of a song or a tune\cite{savageSequence2022a}. 
  However the conflation of the two meanings is tradition-dependent
  rather than a universal music-theoretic fact \cite{ricyi80}. 
  There are cases where the distinction can be inverted when the
  ornamentation technique itself carries the music's identity
  \cite{huangShe2013,pearsonCoarticulation2016}.

  A clean case of the two senses coexisting without confusion is
  Irish traditional music, where performers and teachers use
  ``ornamentation'' specifically for a (largely rhythmic)
  repertoire of techniques that sit on top of the melody and are
  personal to the player\cite{cowderyMelodic1990,osuilleabhainCreative1990}.
  A teacher may pass on their style, but the ornamentation
  is not considered an integral part of the melody:
  the melody is taught and remembered as a separate
  object, while ornamentation belongs to style and to the
  individual.

  \subsection*{Scales for non-melodic pitch}
  Scales are clearly an organizing principle for melodies, but
  pitch is also produced harmonically and percussively, on tuned
  instruments used primarily for rhythm. In Western music
  melody and harmony use the same scales (albeit in different
  ways). Is this always the case?

  In practice, the answer appears to be almost always yes. From the
  standpoint of composition, using a single scale for both
  melody and harmony is vastly more efficient than maintaining
  separate pitch systems. From the standpoint of performance,
  it reduces cognitive load: a singer or instrumentalist
  operating in one scale system does not have to keep in mind a
  parallel set of pitch categories for the accompaniment.
  And from the standpoint of instrument design, a single scale
  means a single set of pitches to be built into or tuned on
  the instrument. Exceptions exist at the level of subsets --
  drone polyphony may use only one or two pitches from the full
  scale, and simple bass lines may omit certain degrees -- but these
  are subsets of a shared system, not independent systems.

  Three exceptions highlight ways that scales can diverge within
  performances. In Georgian three-part singing, \citet{scherbaumIntonation2023}
  showed a small but significant difference in the size of one
  melodic ($\sim$\SI{170}{cents}) and the corresponding
  harmonic interval ($\sim$\SI{200}{cents}) -- the melodic interval
  conforms with the equidistant scale structure, while the harmonic
  interval arises during simultaneous harmonizing with fourths and fifths.
  \citet{diazSinging2024} compared the tuning of the berimbau
  -- a single-string chordophone deployed primarily for rhythm -- to
  the vocalists, finding that sometimes they were tuned together,
  but not always. 
  Finally there are examples of bi- or poly-tonality in early
  twentieth-century Western art music\cite{persichettiTwentiethcentury1961}.
  These three cases -- a small adjustment, a two-note percussive accompaniment,
  and two overlapping conventional scales sounded simultaneously
  -- are noteworthy, yet appear to be deviations from the norm
  rather than a competing paradigm.

  Even allowing for these qualifications, no documented
  tradition appears to use fundamentally different,
  non-overlapping scales for melody and harmony, though it
  would be interesting to know if such cases exist. This
  convergence has consequences for theories of scale origins:
  a constraint based on melodic considerations propagates to harmonic
  practice if the two share a scale, and vice versa\cite{mcbrideMelody2024}.

  \subsection*{Open questions}
  \nsb{Relation between theoretical and empirical scales}
  For the vast majority of historical research on scales, the work
  has been largely theoretical, with the only empirical input being what
  listeners report hearing. Early measurements did involve technology for
  measuring tunings (\eg, monochords, later tuning forks), but these were
  simply standards for comparison -- at the end, a human had to listen
  to two sounds and decide if they were similar. There are two problems 
  to highlight here. First, humans are not only imprecise, but biased --
  what we hear depends on what we expect to hear\cite{ambrazeviciusScales2004}.
  For example, in \citet{brownMusical2025a} we compared scale measurements
  taken purely by ear with those obtained computationally -- fourths
  and fifths were overcounted by about \SIrange{10}{20}{\%}, while
  tritones were undercounted by about \SI{50}{\%}. Second, theory can
  describe music at different levels of granularity, and the grain
  is not always apparent: music theory that looks prescriptive may in
  fact be a coarse summary of variable practice, but without
  measurements we cannot tell which.

  These problems bear directly on how much we should infer from the
  historical record on scale theory. \citet{far04} is doubtful
  about whether the theoretical tuning systems of Persian music
  are reliably produced in practice, as, ``no Middle
  Eastern musical instrument is capable of producing intervals of such precision''.
  Similar questions have been raised about Turkish\cite{sig77}, Arabic\cite{maram93}, 
  and Indian\cite{jairazbhoyIntonation1963,seris11} music.
  Thai music offers a recent example where theory appears to be
  a loose description of statistical trends, and because we have measurements,
  we can actually see the level of granularity it sits at\cite{garan15}.
  For older traditions, we lack the measurements to make the same judgment.
  It would be an overreaction to ignore the vast history of scholarship on
  scales. Instead, the relation between theoretical and empirical
  scales is something that should be studied, so that we have a
  better idea of how much to trust historical scale theory as
  representative of music as it was played.

  \nsb{Human perception of scales}
  When is a song ``in tune'', ``out of tune in one scale'', or ``in tune
  in a different scale''? The perceptual boundary is likely
  related to interval discrimination thresholds (roughly
  \SIrange{50}{100}{cents})\cite{zaratePitchinterval2012}, but it may vary with
  familiarity, training, and culture\cite{mcdermottMusical2010a}.
  Cross-cultural experiments on scale perception would be valuable
  but remain rare\cite{perlmanExperimental1996}.
  What is perceived as a bigger change: an extra scale degree,
  a missing scale degree, or altered scale degrees?
  Do perceptions differ across cultures, and if so, do the
  differences reflect intonation precision in the music itself, or
  something more fundamental about categorical pitch perception?
  Anecdotal examples exist in the literature\cite{kubam80,anija82},
  but the question deserves systematic study.
  
  \nsb{From pitch sets to tonal hierarchies}
  Scales are a compact description of the pitches used in melodies,
  but apart from having a labelled tonic, they do not tell us
  anything about the tonal hierarchy. Understanding how pitch
  is organized by humans in music would benefit from a more
  detailed picture -- one that is provided by the full pitch histogram.
  Considering the weight and variance of the components
  of a fitted GMM, rather than just the means, informs us about
  how regular and how salient each pitch class is.
  Working at this level also softens the binary in/out problem
  -- whether a rare pitch counts as a scale degree is less
  pressing when the quantity of interest is its salience and
  regularity. The cross-cultural study of tonal hierarchies
  is a natural step up from the study of scales.

  \nsb{Joint transcription and scale inference}
  Recovering a scale from audio runs through a chain of
  operations\cite{scherbaumTonal2020,yuComputational2025},
  -- source separation, pitch tracking, onset/offset annotation,
  `melodic pitch' annotation, drift correction, and mixture model
  fitting -- and competent tools exist for each.
  Computers aid reproducibility and precision, but a great deal
  of manual annotation and parameter tuning remains. A fully
  end-to-end automated pipeline is unlikely in the near future,
  except perhaps for genres of Western music (\eg, popular or
  classical) that are overrepresented in the training data used
  in music information retrieval. A more feasible medium-term
  goal is an end-to-end pipeline in which all fine-tuning and
  annotation decisions are fully documented. Such documentation
  of human choices is essential for developing algorithms that
  can match human-level intuition for parsing music from diverse
  sources.

\section*{Conclusion} 
 
  A scale exists in three forms: the physical signal itself, a person's
  expectation/perception, and the cultural convention.
  The empirical core proposed here primarily deals with the signal,
  but is moderated by the other two -- ultimately the three are
  inextricable. Other forms of scales are not dismissed, but rather
  situated in relation to the emprical definition,
  where they can be studied as the cognitive and cultural overlays they are.
  This separation is necessary both for understanding how the forms link,
  and for tackling large-scale cross-cultural analyses in a consistent and
  robust manner. Stripping scales down to their empirical core is reductive,
  and is only one part of the systematic study of music, but it is required
  for studying the organization of pitch across cultures -- especially for
  archival recordings without insider analysts, or in cases where a
  tradition's verbal categories are sparse or absent.

\subsection*{Data and Code Availability}
  Data and code is available at
  \href{github.com/jomimc/ScaleDefinition}{github.com/jomimc/ScaleDefinition}.

\subsection*{AI Usage Declaration}
  Claude was used for analysis code and for copyediting the manuscript.
  The author assumes full responsibility for both.

\bibliography{ScaleDefinition}
\bibliographystyle{unsrtnat}

\clearpage\includepdf[pages=-]{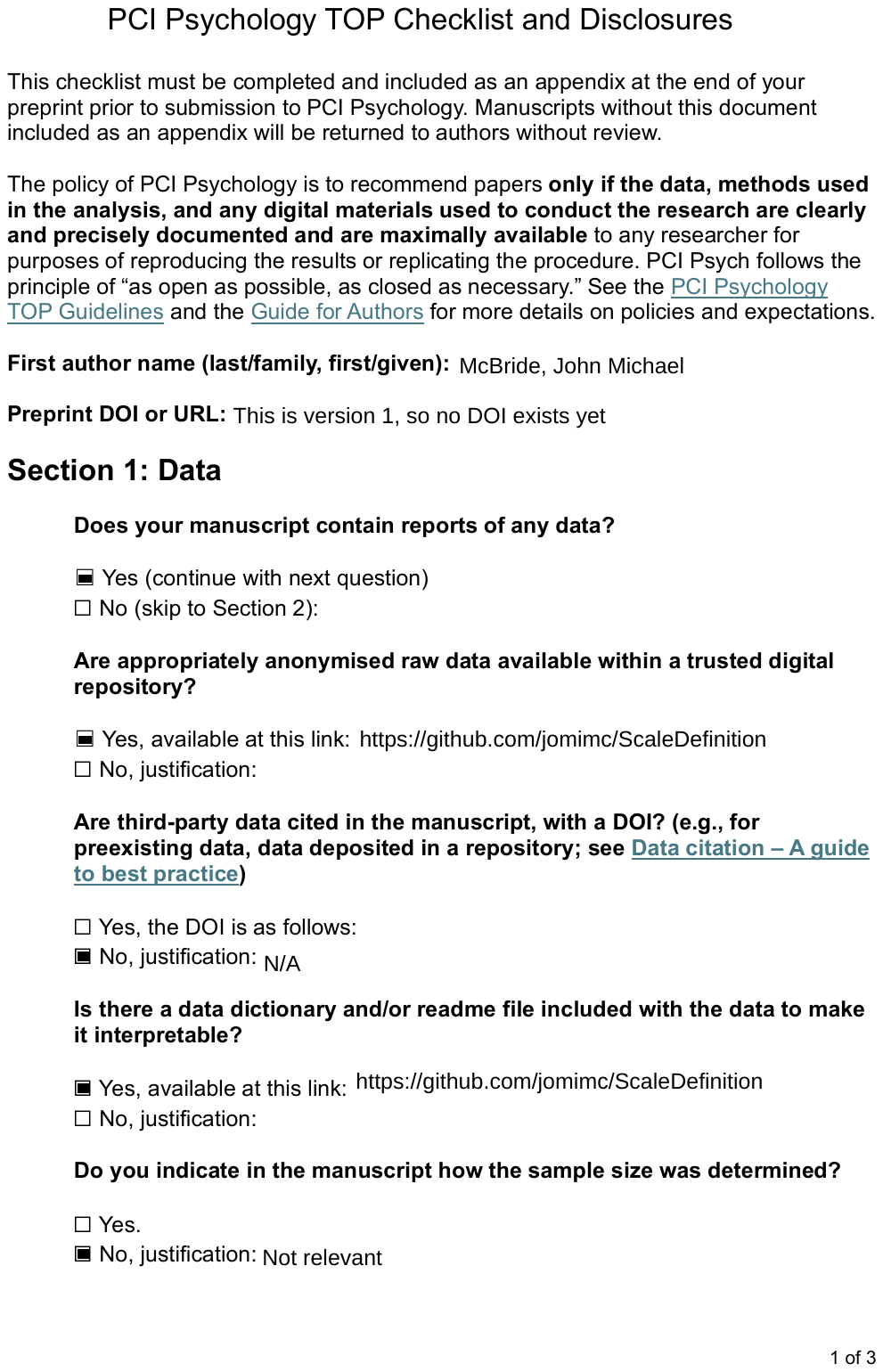}

\end{document}